\documentclass[sn-nature,iicol]{sn-jnl}

\usepackage{graphicx}%
\usepackage{multirow}%
\usepackage{amsmath,amssymb,amsfonts}%
\usepackage{amsthm}%
\usepackage{mathrsfs}%
\usepackage[title]{appendix}%
\usepackage{xcolor}%
\usepackage{textcomp}%
\usepackage{manyfoot}%
\usepackage{booktabs}%
\usepackage{algorithm}%
\usepackage{algorithmicx}%
\usepackage{algpseudocode}%
\usepackage{listings}%
\usepackage{braket}%
\usepackage{verbatim}%
\usepackage{subfigure}

\begin{document}

\title[Article Title]{Generative modeling assisted simulation of measurement-altered quantum criticality}


\author[1]{\fnm{Yuchen} \sur{Zhu}}\email{yzhu738@gatech.edu}

\author*[1]{\fnm{Molei} \sur{Tao}}\email{mtao@gatech.edu}

\author[2]{\fnm{Yuebo} \sur{Jin}}\email{yuebojin@brandeis.edu}

\author*[3]{\fnm{Xie} \sur{Chen}}\email{xiechen@caltech.edu}

\affil[1]{Georgia Institute of Technology}

\affil[2]{Brandeis University}

\affil[3]{California Institute of Technology}


\abstract{
In quantum many-body systems, measurements can induce qualitative new features, but their simulation is hindered by the exponential complexity involved in sampling the measurement results. We propose to use machine learning to assist the simulation of measurement-induced quantum phenomena. In particular, we focus on the measurement-altered quantum criticality protocol and generate local reduced density matrices of the critical chain given random measurement results. Such generation is enabled by a physics-preserving conditional diffusion generative model, which learns an observation-indexed probability distribution of an ensemble of quantum states, and then samples from that distribution given an observation.
}





\maketitle

\section{Introduction}\label{intro}

In quantum mechanics, measurement plays a magic role. Instead of simply reading the information in a state, quantum measurement actively changes the quantum state by collapsing the wave function into one of the measurement basis states, inducing intrinsic randomness that is not available in a classical setting. This unique feature underlies the success of various quantum protocols, like quantum cryptography, quantum games and quantum error correction codes \cite{nielsen2010quantum}. Recently, measurement in many-body quantum systems has received significant attention where it was found to qualitatively alter the universal properties of the system. On the one hand, measurement introduces randomness and decoherence into the otherwise unitary quantum processes and can change qualitatively the pattern of energy, charge, and information propagation through the system. On the other hand, the capability of existing quantum hardware for implementing both unitary and non-unitary dynamics has expanded significantly, making the direct quantum simulation of such protocols a near-term possibility. A protocol that has been extensively studied is the measurement-induced dynamical phase transition \cite{Li2018,Skinner2019,Li2019} where adding stochastic measurement steps into a unitary evolution process can drive the system from a thermal, ergodic phase to a non-thermal, localized phase. The protocol we focus on in this paper is the measurement-altered criticality \cite{Jian2020,Lee2023, Murciano2023,Yang2023,Weinstein2023}, where applying measurements to a critical quantum system can lead to scaling behavior not available in equilibrium systems. 

However, measurement in a many-body quantum setting can also be a curse. A major complication in implementing such protocols on quantum hardware comes directly from the randomness of the measurement results. The randomness applies both to the measurement operations used to induce changes in the many-body systems and to those used to analyze the induced changes. To study the effect of measurement-induced changes, we need to have multiple copies of the same measurement-induced quantum state, but each measurement outcome occurs only with an exponentially small probability (when the number of measurements performed is large). This is usually referred to as the post-selection problem in the study of measurement-induced phenomena. Methods have been proposed to address this problem, but they require efficient classical simulation of the quantum protocol \cite{Lee2022,Garratt2023,Garratt2024}. When the measurement-induced quantum process can be efficiently simulated classically, the cross-correlation between the classical simulation data and quantum simulation data can be used to directly identify feature of the system and surpass the post-selection problem. The more interesting case is, however, when no efficient classical simulation is possible. Whether it is possible to efficiently extract measurement-induced changes in the proposed quantum protocols is an open problem. 

We propose to use generative modeling algorithms to help generate sampling data in the simulation of measurement-induced phenomena. Generative modeling, a specialized subfield of machine learning, is dedicated to creating models capable of generating new data points that closely match the statistical distribution of a given dataset. These models adeptly capture and replicate the patterns, structures, and characteristics of the input data, allowing them to yield new instances that seem to originate from the same distribution. Among them, diffusion model \citep{sohl2015deep,ho2020denoising,song2020score} has risen to prominence for its remarkable ability to produce high-quality data, especially notable in image generation \cite{rombach2022high,ho2022cascaded}. The fundamental mechanism of diffusion model starts with adding noise to the data and progressively altering it into a tractable Gaussian prior with analytically available transition kernels. Meanwhile, how the data distribution gradually gets pushed forward to the prior is learned and encoded in a score function. The model then effectively reverses this noise addition, reconstructing the data from its noised state through the application of the learned score function \citep{anderson1982reverse,haussmann1986time}. 
Diffusion model has demonstrated impressive capabilities in generating various types of dataset across different application area, such as video synthesis \citep{ge2023preserve,blattmann2023align}, language modeling \citep{austin2021structured,lou2023discrete} and point cloud generation \citep{zeng2022lion,luo2021diffusion}. With the recent rise of AI4Science, the potential of diffusion model in tackling scientific problems also started being explored, including applications in biology \citep{jumper2021highly,watson2022broadly}, chemistry \citep{duan2023accurate,hoogeboom2022equivariant}, and climate science \cite{mardani2024residual}. Other important notions relevant to this article include constrained generations \citep[e.g.,][]{liu2023mirror,
fishman2023metropolis,
fishman2023diffusion,
lou2023reflected} and conditional generations \citep[e.g.,][]{dhariwal2021diffusion,ho2022classifier}.

For the purpose of assisting with the simulation of measurement-induced phenomena, we need to generate new samples of quantum states labeled by random measurement results, which is, however, a nontrivial task. Quantum (mixed) states are represented as complex-valued density matrices \cite{nielsen2010quantum}, that have to be Hermitian, positive semi-definite, and trace one. Data-based generation of quantum density matrices therefore amounts to generative modeling in non-Euclidean spaces. Generic generative modeling approaches will almost surely violate the constraints due to indispensable errors originating from finite data and compute. Ref.~\cite{zhu2024quantum} proposed the first machine learning methodology that generates density matrices similar to training data and strictly satisfies their structural constraints, based on hard-wiring these constraints into diffusion models via technique from convex optimization. Combined with the locality property expected in physical quantum many-body systems, the mirror diffusion based generative modeling proposed in Ref.~\cite{zhu2024quantum} provides the perfect tool for helping with the post-selection problem in measurement-induced protocols.

This paper is organized as follows: in section~\ref{sec:MAIC}, we review the measurement protocol we focus on in this paper -- the measurement-altered Ising criticality proposed in Ref.~\cite{Murciano2023}. More details on how we classically simulate the process to generate training data is discussed in Appendix~\ref{ap:simulation}. In section~\ref{sec:MAIC+QSMD}, we discuss how to use the generative modeling method to learn and generate new sampling data of quantum states in the measurement-altered criticality protocol. In particular, we highlight the role played by locality in the quantum system, which makes the generative method we use both essential and powerful for this class of problem.

\section{Measurement-altered Ising quantum critically}\label{sec:MAIC}

In this section, we discuss in detail the measurement-altered quantum criticality protocol \cite{Murciano2023} we aim to simulate and how generative modeling can learn and generate more data based on limited sampling data generated by a quantum simulator.

\begin{figure}[th]
    \centering
    \includegraphics[scale=0.37]{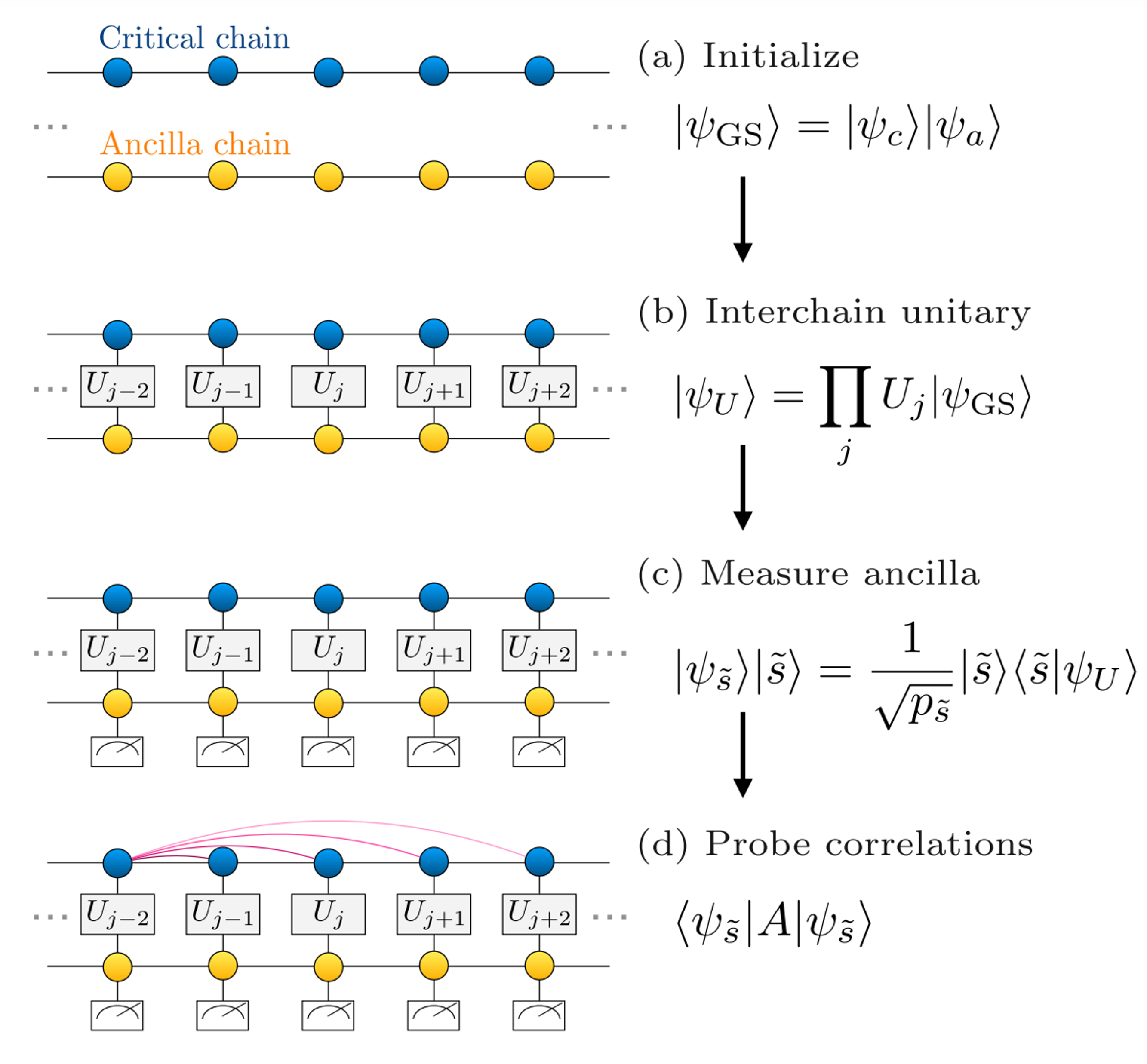}
    \caption{This figure is taken from Fig. 1 of Ref.\cite{Murciano2023} which illustrates the protocol of measurement-altered criticality.}
    \label{fig:MAIC}
\end{figure}

Ref. \cite{Murciano2023} considers how measurement can modify universal properties of a one-dimensional critical Ising chain. Consider a one-dimensional chain of $N$ qubits labeled $1,...,N$, as shown in Fig.~\ref{fig:MAIC}. In terms of Pauli operators,
\begin{align*}
    I = \begin{pmatrix} 1 & 0 \\ 0 & 1 \end{pmatrix},
X = \begin{pmatrix} 0 & 1 \\ 1 & 0 \end{pmatrix},
Z = \begin{pmatrix} 1 & 0 \\ 0 & -1 \end{pmatrix},
Y = \begin{pmatrix} 0 & -i \\ i & 0 \end{pmatrix}.
\end{align*}
the Hamiltonian of the one-dimensional critical Ising chain is given as,
\begin{align}
    H = - \sum_{i=1}^{N} Z_{i} Z_{i + 1} - \sum_{i=1}^{N} X_i
\end{align}
where $\sum_{i}$ is taken to have periodic boundary condition, i.e., $Z_{N}Z_{N+1}$ stands for $Z_{N}Z_{1}$ and is part of the Hamiltonian $H$. Here, $X_i$ means applying Pauli operator $X$ to site $i$ and $Z_iZ_j$ means applying the Pauli operator $Z$ to both site $i$ and site $j$. In short, it holds that,
\begin{align*}
& X_{i} = \underbrace{I \otimes \dots \otimes I}_{\text{Site } 1,\dots, i - 1 } \otimes \underbrace{X}_{\text{Site } i} \otimes \underbrace{I \otimes \dots \otimes I}_{\text{Site } i+1, \dots, N}, \\
& Z_{i}Z_{i + 1} = \underbrace{I \otimes \dots \otimes I}_{\text{Site } 1,\dots, i - 1 } \otimes \underbrace{Z \otimes Z}_{\text{Site }i, i + 1}\otimes \underbrace{I \otimes \dots \otimes I}_{\text{Site } i+2, \dots, N}.
\end{align*}
The ground state (lowest energy state) of $H$ is the so-called critical Ising state, which we denote as $\ket{\psi_{c}}$. The ground state of $H$ is known to have several interesting scaling properties of its correlation function. For example, it satisfies,
\begin{align}
    &\braket{\psi_{c}|Z_i|\psi_{c}} = 0, \, \braket{\psi_{c}|X_i|\psi_{c}} = \frac{2}{\pi}, \, \nonumber\\ 
    &\braket{\psi_{c}|Z_i Z_{i + n}|\psi_{c}} \sim \frac{1}{n^{1/4}}, \, \nonumber \\ 
    &\braket{\psi_{c}|X_i X_{i + n}| \psi_{c}} - \braket{\psi_{c}|X_i|\psi_{c}}\braket{\psi_{c}|X_{i + n}|\psi_{c}} \sim \frac{1}{n^{2}}. \nonumber
\end{align}
Moreover, the entanglement entropy of a segment of length $n$ scales as
\begin{align}
S_n = \frac{1}{3} \log n + c
\end{align}
where $c$ is a constant. The scaling of the correlation functions and the entanglement entropy become more accurate as the length of the chain $N$ and the separation $n$ become larger.

To modify the critical behavior of the chain without completely destroying it, a separate chain called the ancilla chain is introduced. The ancilla chain contains $N$ qubit in a state denoted $|\psi_a\rangle$. The following protocol is used to influence the behavior of the critical chain.

\begin{itemize}
    \item \textbf{Initialization}. 
    The quantum state of the whole system (including both the critical chain as well as the ancilla chain) is given by $\ket{\psi_{GS}} = \ket{\psi_{c}} \otimes \ket{\psi_{a}}$, which is a vector in $\mathbb{C}^{2^{2N}}$.
    \item \textbf{Inter-chain interaction}. A sequence of unitary transformations is applied to the $\ket{\psi_{GS}}$ to obtain an entangled state $\ket{\psi_{U}}$,
    \begin{align}
        \ket{\psi_{U}} = \prod_{j = 1}^{N} U_{j} \ket{\psi_{GS}}
    \end{align}
    where each unitary transformation $U_{j}$ only involves site $j$ on the critical chain and site $j$ on the ancilla chain.
    \item \textbf{Measuring the ancilla chain}. The $N$ qubits in the ancilla chain are measured in some Pauli basis ($X$ or $Z$). The state of the critical chain and the ancilla chain collapses into
    \begin{align}
    |\psi_s\rangle \otimes_i |s_i\rangle, \ s_i = \pm 1, i = 1,...,N
    \end{align}
    \item \textbf{Probe altered criticality}. After obtaining the state $\ket{\psi_s}$ of the critical chain, various measures can be applied to investigate the changes in its critical behavior, such as entanglement or correlation functions.
    
\end{itemize}

It was found in Ref.~\cite{Murciano2023} that this procedure can lead to qualitative and interesting changes to the universal behavior in the critical chain. For example, the scaling of the one-body and two-body correlation functions can change continuously with increasing coupling between the ancilla and the critical chain. This is a very surprising feature that cannot 
appear without the measurement-altered setting. The scaling function of one-body and two-body functions at quantum critical points are highly constrained and need to satisfy very complicated relations. For example, in the case of the Ising critical point, the $1/n^{1/4}$ scaling of the two-body correlation function in the original setting is a universal constant that cannot be changed. Measurement, on the other hand, has the ability to change the scaling and to change it continuously. This observation demonstrates the power of measurements to induce novel physical phenomena that would otherwise be forbidden. 
\section{Generative modeling assisted simulation}
\label{sec:MAIC+QSMD}

\subsection{Effects of locality}
If a reliable quantum simulator is available and we can run the protocol described in Sec.\ref{sec:MAIC} as many times as we want, then for each measurement result $s$, it is in principle possible to reconstruct the corresponding $|\psi_s\rangle$ using quantum state tomography. However, this procedure is computationally very hard. Each $s$ occurs with an exponentially small probability. Sampling through all $s$ takes an exponentially long time that scales as $O(2^N)$. Moreover, given a fixed $s$, it takes an exponentially long time to do a full state tomography on $|\psi_s\rangle$ which leads to another factor of $O(2^N)$ in complexity. Altogether, this makes the problem double-exponentially hard. 

The task can be simplified using a key feature in quantum many-body systems -- locality. First, many important features of the critical state can be extracted from local reduced density matrices (RDM) of a few qubits without knowing the global wavefunction. Therefore, instead of gathering data about $\{s,|\psi_s\rangle\}$, we only need to gather data about $\{s,\rho^i_s\}$, where $\rho^i_s$ are RDMs on a few qubits near site $i$.  
Secondly, because qubits in the critical chain interact only with some of their neighbors throughout the protocol, $\rho^i_s$ is affected mostly by the measurement result of ancilla qubits near site $i$. Therefore, instead of using the full length of $s$ to label $\rho^i$, we can truncate $s$ to a short segment around site $i$, denoted as $s_{[i]}$. 

\begin{figure*}[ht]
    \centering   \includegraphics[height=1.8in]{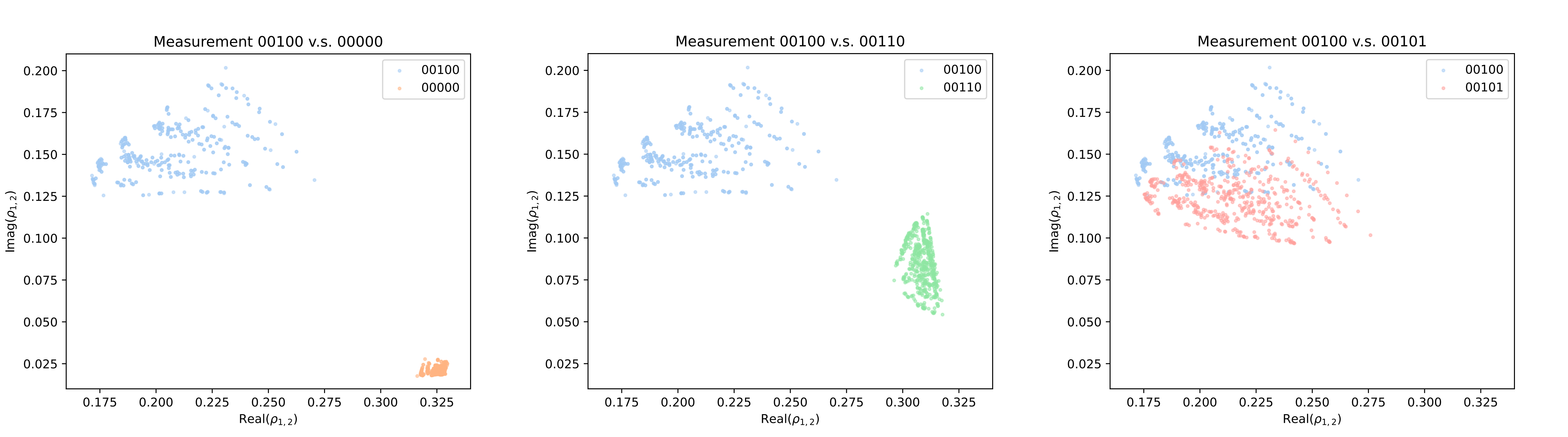}
    \caption{Scattered plots of the imaginary vs the real part of the off-diagonal entry of the one-body reduced density matrix on site $i$ labeled by the measurement result on site $i-2$ to site $i+2$. The difference in distribution decreases as the difference in the measurement result moves away from site $i$.}
    \label{fig:rdm_scatter}
\end{figure*}

To verify the validity of this simplification, we generated the simulation data of a critical chain of length $N=14$ (as described in Appendix~\ref{ap:simulation}) and studied its dependence on different $s_i$'s. In Fig.~\ref{fig:rdm_scatter}, we plot the distribution of the real and imaginary part of the off-diagonal entry of the one-body RDMs on site $i$ labeled by the truncated measurement result on the five sites surrounding it (site $i-2,i-1,i,i+1,i+2$). Among the three plots, we see that the difference between the distribution is the largest when $s_i$ is different while the other parts of the truncated measurement result are the same (e.g. between $00000$ and $00100$). The difference becomes smaller when $s_{i+1}$ is different (e.g. between $00110$ and $00100$) and is the smallest when $s_{i+2}$ is different (e.g. between $00101$ and $00100$). Here, the zero-one sequence refers to the measurement outcome observed at site $i-2$ to site $i+2$, where $0$ stands for observing $\ket{0}$ and $1$ stands for observing $\ket{1}$. 

\begin{figure}[ht]
    \centering       \includegraphics[height=2.2in]{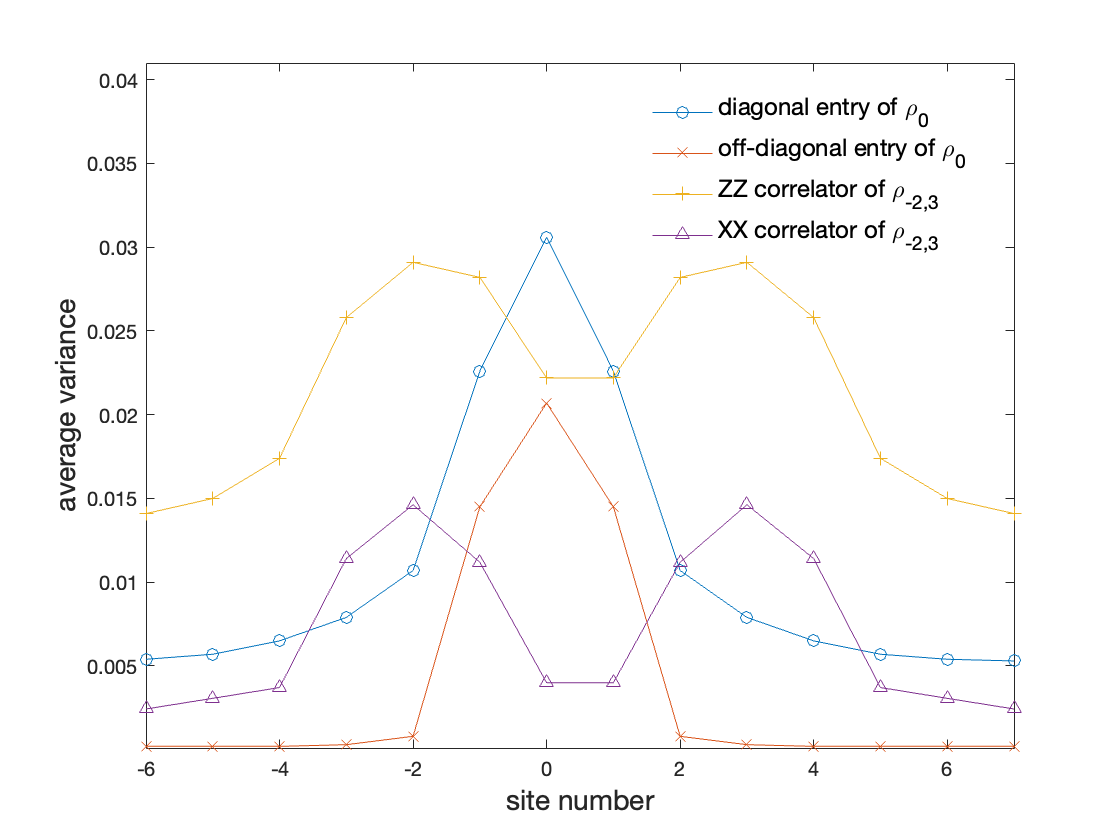}
    \caption{Average variance of one and two-body RDM with respect to the change in measurement results at each site. The blue and red lines represent the diagonal and off-diagonal entries respectively of the one-body RDM at site $0$. The yellow and purple lines represent the $ZZ$ and $XX$ correlators of the two-body RDM at site $-2$ and $3$. We have rescaled the $ZZ$ and $XX$ correlators by a factor of $3$ and $2000$ respectively.}
    \label{fig:variance} 
\end{figure}

In Fig.~\ref{fig:variance}, we plot the variance of one-body and two-body RDMs in $\ket{\psi_s}$ with respect to the change of different $s_i$'s. We label the $14$ sites of the chain as sites $-6, -5, ..., 6,7$. We choose to calculate the one-body RDM on site $0$ and the two-body RDM on site $-2$ and $3$. 
Fixing all but one $s_i$'s, we calculate the variance of the difference entries in the one and two-body reduced density matrices with respect to the change in a particular $s_{i_0}$ and then averaged over all the fixed $s_i$'s. In Fig.~\ref{fig:variance}, we plot the average variance calculated in this way for the diagonal and off-diagonal entries of the one-body RDM, as well as the $ZZ$ and $XX$ correlators of the two-body reduced density matrix (i.e. $\langle Z_{-2}Z_3 \rangle-\langle Z_{-2} \rangle \langle Z_3 \rangle$ and $\langle X_{-2}X_3 \rangle-\langle X_{-2} \rangle \langle X_3 \rangle$).

As Fig.~\ref{fig:variance} shows, the average variance is large with respect to changes in measurement results at sites near the location of the reduced density matrices (i.e. site $0$ for the one-body RDM and site $-2$ and $3$ for the two-body RDM) while the variance decays once we move away from these sites. This is consistent with the result shown in Fig.~\ref{fig:rdm_scatter}. Therefore, most of the variation in the RDMs comes from changes of $s_i$ at its surrounding locations. This is expected due to the locality property of the simulated spin chains. 

The locality feature can help simplify the sampling task. Instead of labeling the post-measurement state with the full measurement result of $s$, we can label them with the truncated $s_{[i]}$. The chance of finding each $s_{[i]}$ in the measurement result will be significantly higher than that of each $s$. Another advantage of using $s_{[i]}$ is that it allows us to extend to a bigger total system size without requiring samples of a bigger system. A complication associated with the truncation of $s$ to $s_{[i]}$ is that, due to the existence of power law correlation in the critical state, $s_{[i]}$ does not uniquely determine $\rho^i$. The shorter the length of $s_{[i]}$, the larger the variance in $\rho_i$. Therefore, while each $s$ uniquely determines $\rho^i$, given the truncated $s_{[i]}$, we have a distribution of $\rho^i$, which we denote as $D(\rho^i_{s_{[i]}})$.

\subsection{Generative modeling of ensembles of quantum states}
In the following, to simplify notation and clearly present the problem set-up, we take $i = 1$ without loss of generality as the system is translation-invariant. Thus, we drop the superscript $i$ for RDMs and use $\rho_{s_{[i]}}$ to stand for the random variable with label $s_{[i]}$, and we denotes distribution of such random variable as $D(\rho_{s_{[i]}})$, and we use $\rho^{j}_{s_{[i]}}, j = 1, \dots, M$ to stand for data points sampled from $D(\rho_{s_{[i]}})$. 

As mentioned before, while the introduction of locality greatly simplified the problem, it brings one additional challenge because the mapping between observations and RDMs is no longer injective, resulting in the RDMs associated with each $s_{[i]}$ being a probability distribution with multiple possible values. This implies that the task of modeling $\rho_{s[i]}$ now involves generative modeling, which is a popular machine learning task. Generative modeling is a problem where one has samples of data and aims at generating more similar samples (i.e. samples from the data distribution), which exactly fits our scenario. Suppose that running the protocol on a quantum simulator a certain number of times generates sample data 
$\rho^j_{s_{[i]}}, j = 1, \dots, M$ for some $s_{[i]}$, then we can learn the probability distribution of the data $D(\rho_{s{[i]}})$ and generate more samples from this distribution. The Structure-Preserving Diffusion Models (SPDM) in \cite{zhu2024quantum} provided a first way for doing so while strictly ensuring that the generated samples are physical (more precisely, satisfying all constraints a density matrix needs to satisfy and thus being quantum states). Moreover, note the task is actually a conditional generation task, because $s_{[i]}$ is the label/condition and for each $s_{[i]}$ there is a probability distribution of data. SPDM had the conditional generation machinery built-in as well, and it can even leverage the extrapolation capability of neural network to generate samples from $D(\rho_{s{[i]}})$ even when $s_{[i]}$ is a new label that's not seen in the training data.

Note that, due to the truncating $s$ into $s_{[i]}$, it is essential that we learn and sample from the reduced density matrices $\rho$, rather than the measurement result of $\rho$ in terms of certain operators. For each $\rho$, when we measure certain operator $O$, we obtain random measurement results which average to $\text{Tr}(\rho O)$. If we have a fixed $\rho$, then from the random measurement results of a proper set of operators, we can reconstruct $\rho$ using quantum state tomography and obtain any other quantity associated with $\rho$. However, if we have a distribution of $\rho$ and sample their collective measurement outcome, we will not be able to distinguish between contributions from different $\rho$'s and reconstruct the collection of $\rho$. Because of that, we will lose access to new quantities that one might want to evaluate using the sampling data, especially those involving nonlinear functions of $\rho$ like entanglement.

\section{Summary and future direction}

In this paper, we propose to employ generative modeling to help with the post-selection problem in simulating measurement-induced phenomena in quantum many-body systems. In particular, we focus on the measurement-altered quantum criticality protocol proposed in Ref.~\cite{Murciano2023}. We found that due to the locality property of the quantum chain under study, the simulation can be reduced to a problem of sampling from distributions of local quantum-reduced density matrices labeled by local measurement results. The mirror-diffusion based quantum state generation method developed in Ref.~\cite{zhu2024quantum} is well suited to this task, by learning from training data generated by a quantum computer/simulator and then generating more sampling data from the learned distribution. We plan to implement that method to demonstrate the effectiveness of the machine learning methodology of generative modeling in assisting with quantum simulation protocols.

\begin{appendices}

\section{Simulation procedure: details}
\label{ap:simulation}
We consider a specific setup of measurement-altered criticality in this paper, namely the case II in Ref. \cite{Murciano2023}.
\begin{itemize}
    \item \textbf{Initialization: } We choose $\ket{\psi_{a}}$ to be a paramagnetic state, which is derived by with the following procedure. We first initialize a state that is the tensor product of $N$ $\ket{+}$: 
    \begin{align*}
        \ket{\psi} = \underbrace{\ket{+} \otimes \dots \otimes \ket{+}}_{\text{N times}}
    \end{align*}
    We perform imaginary time evolution $e^{-k \Delta t H}$ to $\ket{\psi}$ to achieve the paramagnetic state for a small time $\Delta t$ and a small integer $k$. Due to the complexity of computing exactly the operator $e^{-\Delta t H}$, we use Trotter's approximation of the operator,
    \begin{align*}
        e^{-k \Delta t H} \ket{\psi} & = \Big(\prod_{j = 1}^{k} e^{-\Delta t H}\Big) \ket{\psi}\\
        & \approx \prod_{ j = 1}^{k} \Big(\big(e^{\sum_{i} \Delta t X_i}\big) \big(e^{\sum_{i} \Delta t Z_i Z_{i + 1}}\big) \Big) \ket{\psi} \\
        & = \prod_{j = 1}^{k} \Big( \big(\prod_{i} e^{\Delta t X_i} \big) \big( \prod_{i} e^{\Delta t Z_i Z_{i + 1}}\big)\Big) \ket{\psi}
    \end{align*}
    where the second line is an approximations due to $e^{A + B} \approx e^{\Delta t A}e^{\Delta t B}$ when $A$, $B$ do not commute but $\Delta t$ is small, and the last line is an equality due to the fact that $\{X_i\}_{i = 1, \dots, N}$ commute with each other and so do $\{Z_{i} Z_{i + 1}\}_{i = 1, \dots, N}$. Therefore, we initialize $\ket{\psi_a}$ as
    \begin{align*}
        \ket{\psi_{a}} = \prod_{j = 1}^{k} \Big( \big(\prod_{i} e^{\Delta t X_i} \big) \big( \prod_{i} e^{\Delta t Z_i Z_{i + 1}}\big)\Big) \ket{\psi}
    \end{align*}
    Besides $\ket{\psi_a}$, we also need to compute the critical ground state $\ket{\psi_c}$. This can also be done by applying the imaginary time evolution for a long time horizon. Denote $T = K \Delta t$ for a relatively large $K$, and then the ground state satisfies
    \begin{align*}
        \ket{\psi_{c}} & \approx e^{-TH} \ket{\psi} \\
        & \approx \prod_{j = 1}^{K} \Big( \big(\prod_{i} e^{\Delta t X_i} \big) \big( \prod_{i} e^{\Delta t Z_i Z_{i + 1}}\big)\Big) \ket{\psi}
    \end{align*}
    Thus, we iteratively apply $e^{\Delta X_i}$ and $e^{\Delta t Z_i Z_{i+1}}$ to $\ket{\psi}$ for different times to achieve $\ket{\psi_a}$ and $\ket{\psi_c}$. After deriving these two states, the state for the total system can be computed with $\ket{\psi_{GS}} = \ket{\psi_c} \otimes \ket{\psi_{a}}$.

    Note that here $\prod_{i} e^{\Delta t X_i}$ and $ \prod_{i} e^{\Delta t Z_i Z_{i + 1}}$ can be computed efficiently. For $\prod_{i} e^{\Delta t X_i}$, it holds that,
    \begin{align*}
        e^{\Delta t X_i} & = e^{I_{2^{i - 1}} \otimes \Delta t X \otimes I_{2^{N - i}}} \\ & = I_{2^{i - 1}} \otimes e^{\Delta t X} \otimes I_{2^{N - i}} \\
        \prod_{i} e^{\Delta t X_i} & = \prod_{i } I_{2^{i - 1}} \otimes e^{\Delta t X} \otimes I_{2^{N - i}} \\
        & = \underbrace{e^{\Delta t X} \otimes \dots \otimes e^{\Delta t X}}_{\text{N Times}} \\
    \end{align*}
    For $ \prod_{i} e^{\Delta t Z_i Z_{i + 1}}$, note that $Z_iZ_{i+1}$ is a diagonal matrix for each $i$, thus $\sum_{i} \Delta t Z_i Z_{i + 1}$ is also a diagonal matrix. To compute $e^{\sum_i \Delta t Z_i Z_{i + 1}}$, we just need to compute the diagonal of $\sum_i \Delta t Z_i Z_{i + 1}$ and take entry-wise exponential. Therefore, both of the operator can be computed efficiently. 
    \item \textbf{Inter-chain unitary and ancilla chain measurement }. We consider the following form of $U_j$, 
    \begin{align*}
        U_{j} = \exp\Big(i u \big( Z_j - C\big) \tilde X_j\Big)
    \end{align*}
    where $u \ll 1$ is a small positive number, $C = -1$ is a constant. For the measurement operator applied to the ancilla chain, we use $\tilde Z$. Therefore the measurement result of qubit $j$ on the ancilla chain $\ket{\tilde s_j} \in \{\ket{0}, \ket{1}\}$. In fact, the effect of inter-chain unitary transformation and ancilla chain measurement can be combined and summarize it into the following operator applied to $\ket{\psi_c}$ to obtain the post-measurement state of the critical chain,
    \begin{align*}
        \ket{\psi_{\tilde s}} = \frac{1}{\sqrt{\mathcal{N}}} U^{\prime} e^{-H_{m} /2} \ket{\psi_{c}}
    \end{align*}
    where $\mathcal{N}$ is a normalizing constant, $U^{\prime}$ is a unitary operator acting solely on the critical chain, $H_{m}$ is a Hermitian operator that encodes the non-unitary changes imposed by the measurement. In fact, defining $U^{\prime} = e^{i H^{\prime}}$, with an approximation error $O(u^2)$, we have the following formula for $H^{\prime}$ and $H_{m}$,
    \begin{align*}
        H^{\prime} & = u \sum_{j} a(j)Z_{j}, \\
        H_m & = u^2 \sum_{j} m_j Z_{j} + u^{2} \sum_{j \neq k}V_{jk} Z_{j}Z_{k}
    \end{align*}
    where the coefficients $a(j,k), a(j)$ are given by,
    \begin{align*}
        a(j,k) = \frac{\braket{\tilde s|\tilde X_j \tilde X_k|\psi_a}}{\braket{\tilde s|\psi_a}}, \, a(j) = \frac{\braket{\tilde s|\tilde X_j|\psi_a}}{\braket{\tilde s|\psi_a}}
    \end{align*}
    And $m_j, V_{jk}$ are given by,
    \begin{align*}
        V_{jk} = a(j,k) - a(j) a(k), \\
        m_{j} = 2\left(1-a(j)^2+\sum_{k\neq j}V_{jk}\right)
    \end{align*}
    Note that, here $H^{\prime}$ and $H_{m}$ can be computed efficiently since each $Z_j$, $Z_j Z_k$ are all diagonal matrices. Moreover, we can further simplify $e^{iH^{\prime}}$,
    \begin{align*}
        e^{iH^{\prime}} = e^{iua(1) Z} \otimes e^{i u a(2) Z} \otimes \dots \otimes e^{i u a(N) Z}
    \end{align*}
    \item \textbf{Probe correlation function }. After obtaining ancilla chain measurement result $\ket{\tilde s}$ and corresponding critical chain state $\ket{\psi_{\tilde s}}$, we measure the expectation of some single site observable, which requires the qubit $j$ reduced density matrix $\rho_{j}$ for each $1 \leq j \leq N$. The site $j$ RDM can be computed by taking the partial trace to trace out the rest of the qubits. Note that the chain is cyclically permutation-equivariant. Because of this, we should associate each $\rho_{j}$ with a permuted observation outcome $\ket{\tilde s^{j}}$ that starts with site $j$,
    \begin{align*}
        \ket{\tilde s^{j}} = \ket{\tilde s_{j}} \otimes \ket{\tilde s_{j + 1}} \otimes \dots \otimes \ket{\tilde s_{j - 2}} \otimes \ket{\tilde s_{j - 1}}
    \end{align*}

\end{itemize}

\end{appendices}

\bibliography{references}%

\begin{thebibliography}{10}
\expandafter\ifx\csname url\endcsname\relax
  \def\url#1{\burl{#1}}\fi
\expandafter\ifx\csname urlprefix\endcsname\relax\def\urlprefix{URL }\fi
\providecommand{\bibinfo}[2]{#2}
\providecommand{\eprint}[2][]{\url{#2}}
\providecommand{\doi}[1]{\url{https://doi.org/#1}}
\bibcommenthead

\bibitem{nielsen2010quantum}
\bibinfo{author}{Nielsen, M.~A.} \& \bibinfo{author}{Chuang, I.~L.}
\newblock \emph{\bibinfo{title}{Quantum computation and quantum information}}  (\bibinfo{publisher}{Cambridge university press}, \bibinfo{year}{2010}).

\bibitem{Li2018}
\bibinfo{author}{Li, Y.}, \bibinfo{author}{Chen, X.} \& \bibinfo{author}{Fisher, M. P.~A.}
\newblock \bibinfo{title}{Quantum zeno effect and the many-body entanglement transition}.
\newblock \emph{\bibinfo{journal}{Phys. Rev. B}} \textbf{\bibinfo{volume}{98}}, \bibinfo{pages}{205136} (\bibinfo{year}{2018}).
\newblock \urlprefix\url{https://link.aps.org/doi/10.1103/PhysRevB.98.205136}.

\bibitem{Skinner2019}
\bibinfo{author}{Skinner, B.}, \bibinfo{author}{Ruhman, J.} \& \bibinfo{author}{Nahum, A.}
\newblock \bibinfo{title}{Measurement-induced phase transitions in the dynamics of entanglement}.
\newblock \emph{\bibinfo{journal}{Phys. Rev. X}} \textbf{\bibinfo{volume}{9}}, \bibinfo{pages}{031009} (\bibinfo{year}{2019}).
\newblock \urlprefix\url{https://link.aps.org/doi/10.1103/PhysRevX.9.031009}.

\bibitem{Li2019}
\bibinfo{author}{Li, Y.}, \bibinfo{author}{Chen, X.} \& \bibinfo{author}{Fisher, M. P.~A.}
\newblock \bibinfo{title}{Measurement-driven entanglement transition in hybrid quantum circuits}.
\newblock \emph{\bibinfo{journal}{Phys. Rev. B}} \textbf{\bibinfo{volume}{100}}, \bibinfo{pages}{134306} (\bibinfo{year}{2019}).
\newblock \urlprefix\url{https://link.aps.org/doi/10.1103/PhysRevB.100.134306}.

\bibitem{Jian2020}
\bibinfo{author}{Jian, C.-M.}, \bibinfo{author}{You, Y.-Z.}, \bibinfo{author}{Vasseur, R.} \& \bibinfo{author}{Ludwig, A. W.~W.}
\newblock \bibinfo{title}{Measurement-induced criticality in random quantum circuits}.
\newblock \emph{\bibinfo{journal}{Phys. Rev. B}} \textbf{\bibinfo{volume}{101}}, \bibinfo{pages}{104302} (\bibinfo{year}{2020}).
\newblock \urlprefix\url{https://link.aps.org/doi/10.1103/PhysRevB.101.104302}.

\bibitem{Lee2023}
\bibinfo{author}{Lee, J.~Y.}, \bibinfo{author}{Jian, C.-M.} \& \bibinfo{author}{Xu, C.}
\newblock \bibinfo{title}{Quantum criticality under decoherence or weak measurement}.
\newblock \emph{\bibinfo{journal}{PRX Quantum}} \textbf{\bibinfo{volume}{4}}, \bibinfo{pages}{030317} (\bibinfo{year}{2023}).
\newblock \urlprefix\url{https://link.aps.org/doi/10.1103/PRXQuantum.4.030317}.

\bibitem{Murciano2023}
\bibinfo{author}{Murciano, S.}, \bibinfo{author}{Sala, P.}, \bibinfo{author}{Liu, Y.}, \bibinfo{author}{Mong, R. S.~K.} \& \bibinfo{author}{Alicea, J.}
\newblock \bibinfo{title}{Measurement-altered ising quantum criticality}.
\newblock \emph{\bibinfo{journal}{Phys. Rev. X}} \textbf{\bibinfo{volume}{13}}, \bibinfo{pages}{041042} (\bibinfo{year}{2023}).
\newblock \urlprefix\url{https://link.aps.org/doi/10.1103/PhysRevX.13.041042}.

\bibitem{Yang2023}
\bibinfo{author}{Yang, Z.}, \bibinfo{author}{Mao, D.} \& \bibinfo{author}{Jian, C.-M.}
\newblock \bibinfo{title}{Entanglement in a one-dimensional critical state after measurements}.
\newblock \emph{\bibinfo{journal}{Phys. Rev. B}} \textbf{\bibinfo{volume}{108}}, \bibinfo{pages}{165120} (\bibinfo{year}{2023}).
\newblock \urlprefix\url{https://link.aps.org/doi/10.1103/PhysRevB.108.165120}.

\bibitem{Weinstein2023}
\bibinfo{author}{Weinstein, Z.}, \bibinfo{author}{Sajith, R.}, \bibinfo{author}{Altman, E.} \& \bibinfo{author}{Garratt, S.~J.}
\newblock \bibinfo{title}{Nonlocality and entanglement in measured critical quantum ising chains}.
\newblock \emph{\bibinfo{journal}{Phys. Rev. B}} \textbf{\bibinfo{volume}{107}}, \bibinfo{pages}{245132} (\bibinfo{year}{2023}).
\newblock \urlprefix\url{https://link.aps.org/doi/10.1103/PhysRevB.107.245132}.

\bibitem{Lee2022}
\bibinfo{author}{Lee, J.~Y.}, \bibinfo{author}{Ji, W.}, \bibinfo{author}{Bi, Z.} \& \bibinfo{author}{Fisher, M. P.~A.}
\newblock \bibinfo{title}{Decoding measurement-prepared quantum phases and transitions: from ising model to gauge theory, and beyond} (\bibinfo{year}{2022}).
\newblock \urlprefix\url{https://arxiv.org/abs/2208.11699}.
\newblock \eprint{2208.11699}.

\bibitem{Garratt2023}
\bibinfo{author}{Garratt, S.~J.}, \bibinfo{author}{Weinstein, Z.} \& \bibinfo{author}{Altman, E.}
\newblock \bibinfo{title}{Measurements conspire nonlocally to restructure critical quantum states}.
\newblock \emph{\bibinfo{journal}{Phys. Rev. X}} \textbf{\bibinfo{volume}{13}}, \bibinfo{pages}{021026} (\bibinfo{year}{2023}).
\newblock \urlprefix\url{https://link.aps.org/doi/10.1103/PhysRevX.13.021026}.

\bibitem{Garratt2024}
\bibinfo{author}{Garratt, S.~J.} \& \bibinfo{author}{Altman, E.}
\newblock \bibinfo{title}{Probing postmeasurement entanglement without postselection}.
\newblock \emph{\bibinfo{journal}{PRX Quantum}} \textbf{\bibinfo{volume}{5}}, \bibinfo{pages}{030311} (\bibinfo{year}{2024}).
\newblock \urlprefix\url{https://link.aps.org/doi/10.1103/PRXQuantum.5.030311}.

\bibitem{sohl2015deep}
\bibinfo{author}{Sohl-Dickstein, J.}, \bibinfo{author}{Weiss, E.}, \bibinfo{author}{Maheswaranathan, N.} \& \bibinfo{author}{Ganguli, S.}
\newblock \bibinfo{title}{{Deep unsupervised learning using nonequilibrium thermodynamics}}.
\newblock \emph{\bibinfo{journal}{ICML}}  (\bibinfo{year}{2015}).

\bibitem{ho2020denoising}
\bibinfo{author}{Ho, J.}, \bibinfo{author}{Jain, A.} \& \bibinfo{author}{Abbeel, P.}
\newblock \bibinfo{title}{Denoising diffusion probabilistic models}.
\newblock \emph{\bibinfo{journal}{Advances in neural information processing systems}} \textbf{\bibinfo{volume}{33}}, \bibinfo{pages}{6840--6851} (\bibinfo{year}{2020}).

\bibitem{song2020score}
\bibinfo{author}{Song, Y.} \emph{et~al.}
\newblock \bibinfo{title}{Score-based generative modeling through stochastic differential equations}.
\newblock \emph{\bibinfo{journal}{ICLR}}  (\bibinfo{year}{2021}).

\bibitem{rombach2022high}
\bibinfo{author}{Rombach, R.}, \bibinfo{author}{Blattmann, A.}, \bibinfo{author}{Lorenz, D.}, \bibinfo{author}{Esser, P.} \& \bibinfo{author}{Ommer, B.}
\newblock \bibinfo{title}{High-resolution image synthesis with latent diffusion models}.
\newblock \emph{\bibinfo{journal}{IEEE/CVF conference on computer vision and pattern recognition (CVPR)}}  (\bibinfo{year}{2022}).

\bibitem{ho2022cascaded}
\bibinfo{author}{Ho, J.} \emph{et~al.}
\newblock \bibinfo{title}{Cascaded diffusion models for high fidelity image generation}.
\newblock \emph{\bibinfo{journal}{The Journal of Machine Learning Research}} \textbf{\bibinfo{volume}{23}}, \bibinfo{pages}{2249--2281} (\bibinfo{year}{2022}).

\bibitem{anderson1982reverse}
\bibinfo{author}{Anderson, B.~D.}
\newblock \bibinfo{title}{{Reverse-time diffusion equation models}}.
\newblock \emph{\bibinfo{journal}{Stochastic Processes and their Applications}} \textbf{\bibinfo{volume}{12}}, \bibinfo{pages}{313--326} (\bibinfo{year}{1982}).

\bibitem{haussmann1986time}
\bibinfo{author}{Haussmann, U.~G.} \& \bibinfo{author}{Pardoux, E.}
\newblock \bibinfo{title}{Time reversal of diffusions}.
\newblock \emph{\bibinfo{journal}{The Annals of Probability}} \bibinfo{pages}{1188--1205} (\bibinfo{year}{1986}).

\bibitem{ge2023preserve}
\bibinfo{author}{Ge, S.} \emph{et~al.}
\newblock \bibinfo{title}{Preserve your own correlation: A noise prior for video diffusion models}.
\newblock \emph{\bibinfo{journal}{IEEE/CVF International Conference on Computer Vision (ICCV)}}  (\bibinfo{year}{2023}).

\bibitem{blattmann2023align}
\bibinfo{author}{Blattmann, A.} \emph{et~al.}
\newblock \bibinfo{title}{Align your latents: High-resolution video synthesis with latent diffusion models}.
\newblock \emph{\bibinfo{journal}{Proceedings of the IEEE/CVF Conference on Computer Vision and Pattern Recognition}}  (\bibinfo{year}{2023}).

\bibitem{austin2021structured}
\bibinfo{author}{Austin, J.}, \bibinfo{author}{Johnson, D.~D.}, \bibinfo{author}{Ho, J.}, \bibinfo{author}{Tarlow, D.} \& \bibinfo{author}{Van Den~Berg, R.}
\newblock \bibinfo{title}{Structured denoising diffusion models in discrete state-spaces}.
\newblock \emph{\bibinfo{journal}{Advances in Neural Information Processing Systems}} \textbf{\bibinfo{volume}{34}}, \bibinfo{pages}{17981--17993} (\bibinfo{year}{2021}).

\bibitem{lou2023discrete}
\bibinfo{author}{Lou, A.}, \bibinfo{author}{Meng, C.} \& \bibinfo{author}{Ermon, S.}
\newblock \bibinfo{title}{Discrete diffusion language modeling by estimating the ratios of the data distribution}.
\newblock \emph{\bibinfo{journal}{NeurIPS}}  (\bibinfo{year}{2023}).

\bibitem{zeng2022lion}
\bibinfo{author}{Zeng, X.} \emph{et~al.}
\newblock \bibinfo{title}{Lion: Latent point diffusion models for 3d shape generation}.
\newblock \emph{\bibinfo{journal}{arXiv preprint arXiv:2210.06978}}  (\bibinfo{year}{2022}).

\bibitem{luo2021diffusion}
\bibinfo{author}{Luo, S.} \& \bibinfo{author}{Hu, W.}
\newblock \bibinfo{title}{Diffusion probabilistic models for 3d point cloud generation}.
\newblock \emph{\bibinfo{journal}{IEEE/CVF Conference on Computer Vision and Pattern Recognition (CVPR)}}  (\bibinfo{year}{2021}).

\bibitem{jumper2021highly}
\bibinfo{author}{Jumper, J.} \emph{et~al.}
\newblock \bibinfo{title}{Highly accurate protein structure prediction with alphafold}.
\newblock \emph{\bibinfo{journal}{Nature}} \textbf{\bibinfo{volume}{596}}, \bibinfo{pages}{583--589} (\bibinfo{year}{2021}).

\bibitem{watson2022broadly}
\bibinfo{author}{Watson, J.~L.} \emph{et~al.}
\newblock \bibinfo{title}{Broadly applicable and accurate protein design by integrating structure prediction networks and diffusion generative models}.
\newblock \emph{\bibinfo{journal}{BioRxiv}} \bibinfo{pages}{2022--12} (\bibinfo{year}{2022}).

\bibitem{duan2023accurate}
\bibinfo{author}{Duan, C.}, \bibinfo{author}{Du, Y.}, \bibinfo{author}{Jia, H.} \& \bibinfo{author}{Kulik, H.~J.}
\newblock \bibinfo{title}{Accurate transition state generation with an object-aware equivariant elementary reaction diffusion model}.
\newblock \emph{\bibinfo{journal}{Nature Computational Science}}  (\bibinfo{year}{2023}).

\bibitem{hoogeboom2022equivariant}
\bibinfo{author}{Hoogeboom, E.}, \bibinfo{author}{Satorras, V.~G.}, \bibinfo{author}{Vignac, C.} \& \bibinfo{author}{Welling, M.}
\newblock \bibinfo{title}{Equivariant diffusion for molecule generation in 3d}.
\newblock \emph{\bibinfo{journal}{International conference on machine learning (ICML)}}  (\bibinfo{year}{2022}).

\bibitem{mardani2024residual}
\bibinfo{author}{Mardani, M.} \emph{et~al.}
\newblock \bibinfo{title}{Residual diffusion modeling for km-scale atmospheric downscaling}  (\bibinfo{year}{2024}).

\bibitem{liu2023mirror}
\bibinfo{author}{Liu, G.-H.}, \bibinfo{author}{Chen, T.}, \bibinfo{author}{Theodorou, E.~A.} \& \bibinfo{author}{Tao, M.}
\newblock \bibinfo{title}{Mirror diffusion models for constrained and watermarked generation}.
\newblock \emph{\bibinfo{journal}{NeurIPS}}  (\bibinfo{year}{2023}).

\bibitem{fishman2023metropolis}
\bibinfo{author}{Fishman, N.}, \bibinfo{author}{Klarner, L.}, \bibinfo{author}{Mathieu, E.}, \bibinfo{author}{Hutchinson, M.} \& \bibinfo{author}{De~Bortoli, V.}
\newblock \bibinfo{title}{Metropolis sampling for constrained diffusion models}.
\newblock \emph{\bibinfo{journal}{NeurIPS}}  (\bibinfo{year}{2023}).

\bibitem{fishman2023diffusion}
\bibinfo{author}{Fishman, N.}, \bibinfo{author}{Klarner, L.}, \bibinfo{author}{De~Bortoli, V.}, \bibinfo{author}{Mathieu, E.} \& \bibinfo{author}{Hutchinson, M.}
\newblock \bibinfo{title}{Diffusion models for constrained domains}.
\newblock \emph{\bibinfo{journal}{TMLR}}  (\bibinfo{year}{2023}).

\bibitem{lou2023reflected}
\bibinfo{author}{Lou, A.} \& \bibinfo{author}{Ermon, S.}
\newblock \bibinfo{title}{Reflected diffusion models}.
\newblock \emph{\bibinfo{journal}{ICML}}  (\bibinfo{year}{2023}).

\bibitem{dhariwal2021diffusion}
\bibinfo{author}{Dhariwal, P.} \& \bibinfo{author}{Nichol, A.}
\newblock \bibinfo{title}{Diffusion models beat gans on image synthesis}.
\newblock \emph{\bibinfo{journal}{Advances in neural information processing systems}} \textbf{\bibinfo{volume}{34}}, \bibinfo{pages}{8780--8794} (\bibinfo{year}{2021}).

\bibitem{ho2022classifier}
\bibinfo{author}{Ho, J.} \& \bibinfo{author}{Salimans, T.}
\newblock \bibinfo{title}{Classifier-free diffusion guidance}.
\newblock \emph{\bibinfo{journal}{arXiv preprint arXiv:2207.12598}}  (\bibinfo{year}{2022}).

\bibitem{zhu2024quantum}
\bibinfo{author}{Zhu, Y.}, \bibinfo{author}{Chen, T.}, \bibinfo{author}{Theodorou, E.~A.}, \bibinfo{author}{Chen, X.} \& \bibinfo{author}{Tao, M.}
\newblock \bibinfo{title}{Quantum state generation with structure-preserving diffusion model} (\bibinfo{year}{2024}).
\newblock \urlprefix\url{https://arxiv.org/abs/2404.06336}.
\newblock \eprint{2404.06336}.

\end{thebibliography}

\end{document}